# Adaptive Cache-Friendly Priority Queue: Enhancing Heap-Tree Efficiency for Modern Computing

Kiarash Parvizi

Department of Computer Engineering, Bu-Ali Sina University, Hamedan, Iran

## Abstract:

Priority queues are fundamental data structures with widespread applications in various domains, including graph algorithms and network simulations. Their performance critically impacts the overall efficiency of these algorithms. Traditional priority queue implementations often face cache-related performance bottlenecks, especially in modern computing environments with hierarchical memory systems. To address this challenge, we propose an adaptive cache-friendly priority queue that utilizes three adjustable parameters to optimize the heap tree structure for specific system conditions by making a tradeoff between cache friendliness and the average number of cpu instructions needed to carry out the data structure operations. Compared to the implicit binary tree model, our approach significantly reduces the number of cache misses and improves performance, as demonstrated through rigorous testing on the heap sort algorithm. We employ a search method to determine the optimal parameter values, eliminating the need for manual configuration. Furthermore, our data structure is laid out in a single compact block of memory, minimizing the memory consumption and can dynamically grow without the need for costly heap tree reconstructions. The adaptability of our cache-friendly priority queue makes it particularly well-suited for modern computing environments with diverse system architectures.

## Introduction

In the realm of evolving computer memory systems, it is crucial to develop algorithms that align with the intricate structure of memory. Modern memory systems are characterized by a hierarchical arrangement, encompassing cache levels, main memory, and secondary storage. This stands in sharp contrast to conventional computational models, which assumed a uniform 'flat' memory access latency. Nowadays, computing platforms exhibit significant variations in access times across different memory levels, differing by several orders of magnitude. This phenomenon underscores the necessity for algorithms that can navigate this complexity effectively.

Traditionally, the majority of algorithmic research has been conducted within the framework of the Random Access Machine (RAM) model of computation, which assumes a "flat" memory structure with uniform access time. In this standard RAM model, algorithm efficiency is gauged by the number of instructions it entails, with memory access considered to have a unit cost, regardless of the data's location within the memory. However, this model proves inadequate for accurately understanding modern computers equipped with intricate memory hierarchies [2]. The advent of a memory hierarchy implies varying access costs based on the data's placement within the hierarchy. Algorithms that disregard this memory hierarchy face substantial performance penalties.

An alternative approach is the standard two-level memory hierarchy model incorporating block transfers, recognized under different names such as the external memory model, the I/O model, or the disk access model [11]. This model characterizes a computer system with two levels: the cache, located proximally to the CPU, offers swift but limited access, while the disk, situated distantly from the CPU, provides expensive access but offers virtually limitless storage. The difference in access times between these two memory systems and the simplicity of this model render it a compelling choice for analyzing the operational costs of algorithms.

The design of memory hierarchies often leverages the principle of locality of reference to optimize data access. This concept encompasses two key aspects. First, when a program accesses a specific memory location, it is likely to revisit that location multiple times within a short timeframe, known as temporal locality. Second, once a program accesses a particular memory location, it is probable that it will also access nearby memory locations shortly after, termed spatial locality [8].

The fundamental structure of a cache involves multiple cache blocks, each capable of storing multiple words. This design choice is made to leverage spatial locality, as accessing nearby memory locations often occurs in programs. The effectiveness of a cache system lies in its ability to store data that a program is likely to access again. When a program requests data that is not present in the nearest memory level, it results in a cache miss. During a cache miss, the requested word must be retrieved from the next, slower memory level, leading to performance delays. To address this challenge, researchers have explored three fundamental data placement design principles: clustering, coloring, and compression [6, 8]. These principles, when effectively combined, give rise to cache-conscious data structures, which play a pivotal role in optimizing system performance and minimizing access latency. The process of clustering involves organizing data structure elements within a cache block based on the likelihood of them being accessed together, thereby improving spatial and temporal locality [6]. In the context of our heap data structure discussed in this research paper, clustering forms the core design principle. Furthermore, our implementation of the heap structure incorporates the pointer elimination technique by organizing the entire data structure into a single contiguous block of memory.

## Related Work

One early attempt at optimizing cache efficiency in priority queues was conducted by Anthony LaMarca and Richard Ladner in their paper “The influence of caches on the performance of heaps” [2], where they employed a technique involving padding the data array to achieve cache alignment of the memory layout of the implicit heap tree. This strategy ensures that all children of a given node reside within a single cache block, thereby minimizing cache misses. While this approach demonstrated significant reductions in cache misses for very small cache sizes, as outlined in their 1996 paper, its efficacy diminishes with larger computer cache block sizes. This is attributed to the decreasing likelihood of a node having children in two distinct cache blocks. Primarily due to the characteristic of implicit d’ary heaps, which typically exhibit a limited number of children. Moreover, the space required to store this restricted number of children is significantly smaller than the cache block size.

Another approach researchers have proposed for this problem is that instead of utilizing a single tree structure to store priority queue elements, they manage several fully sorted sequences [4,13]. These sequences are subsequently merged to create new sequences as the data structure expands. The key concept is that retrieving elements from a sorted block of memory is significantly more cache-efficient compared to organizing a tree structure in the same block. This is because the former approach involves fewer memory jumps, allowing for linear memory access. The challenge in this approach is to design a data structure that limits the number of sequences it requires and minimizes the average number of jumps between sequences during data extraction. Notable research papers on this topic include Gerth Stølting Brodal and Rolf Fagerberg's work on their cache oblivious design “Funnel Heap” [13], and Peter Sanders exploration of “Fast Priority Queues for Cached Memory” [4].

Sequence heaps have proven to be effective in substantially minimizing cache misses and consequently reducing execution time when executing algorithms on extensive datasets. However, their utility is constrained by increased memory consumption and less favorable worst-case running times for individual priority queue operations, primarily attributable to the sequence merging operations, in other word they only exhibit efficiency gains primarily in an amortized sense. Making these models less suitable for deployment in real-time systems.

One theoretical attempt to formulate a sequence heap model with worst-case efficiency was undertaken by Brodal and Katajainen. In their paper titled “Worst-case efficient external-memory priority queues” [12], they introduced the concept of incremental merging to distribute the workload required for the k-way merging process across multiple calls to the data structure. Regrettably, to the best of our knowledge as of November 2023, there is no publicly available implementation of this method on the internet, and no experimental studies have been conducted to assess the effectiveness of this model.

## Dynamic Cache-Friendly layout:

The designed data structure, named Par-Heap, necessitates the determination of three parameters to specify its schema or memory layout. By adjusting the values of these parameters, various structures and clusters can be generated, endowing the data structure with remarkable flexibility and dynamism. The implementation of the heap is array-based; wherein a heap of size N is accommodated within an array of the same size, enabling the performance of the heap-sort algorithm in-place with only constant extra memory needed.

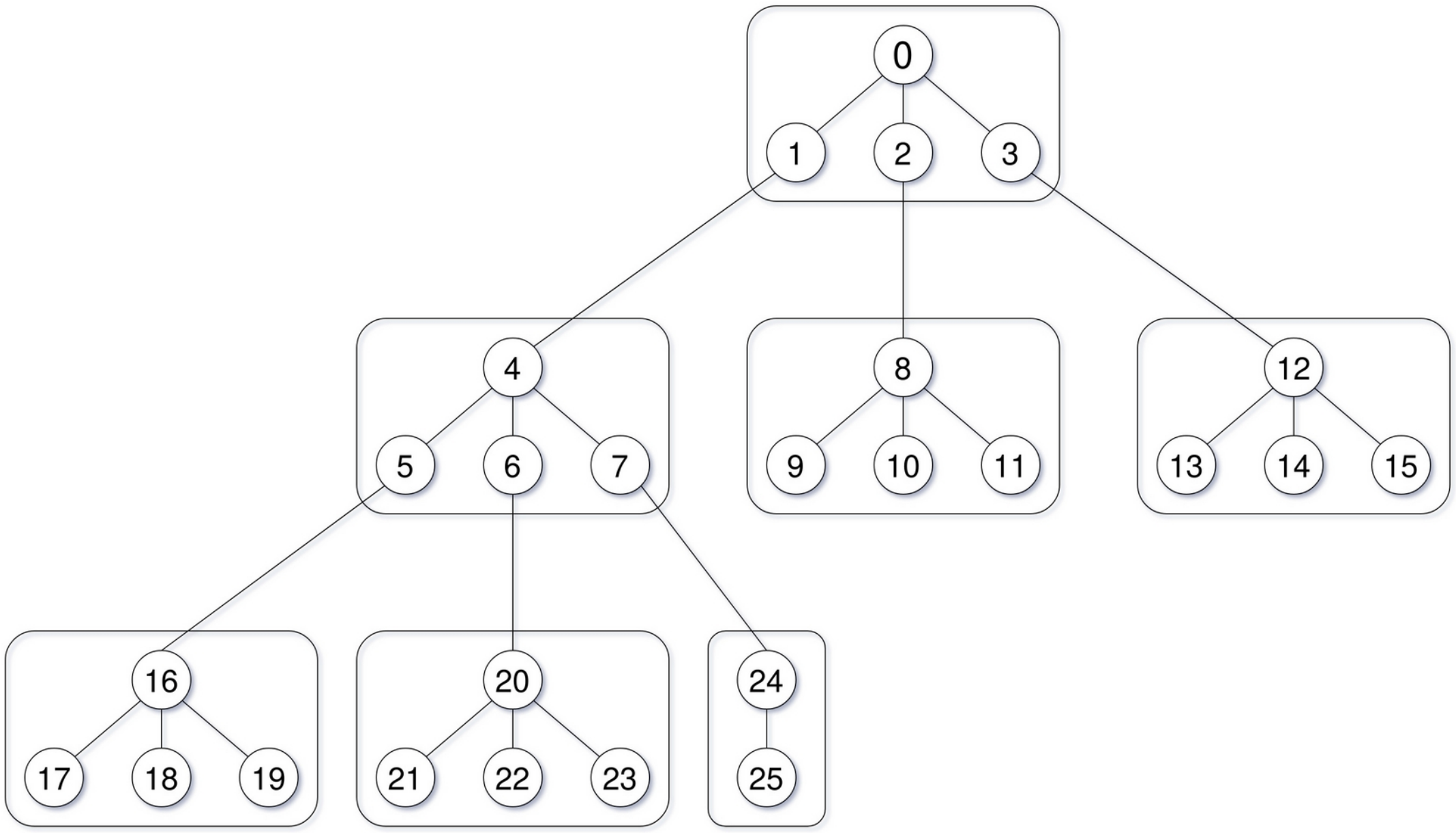

*Figure 1: Par-Heap sample tree memory layout with parameters: intra_child_count = 3, inter_child_count = 1, block_depth = 1*

In Figure 1, it is evident that every block in the structure encompasses a full tree characterized by a depth of 'block_depth' and a fanout of 'intra_child_count.' It is noteworthy that these specific parameters remain consistent across all blocks within the data structure. There is, however, an exception for the final block, serving as the terminal segment of the tree, which necessitates accommodating at least one node. The maximum number of children per block leaf node is determined by the 'inter_child_count' parameter, which maintains a constant value for all block leaf nodes. By placing these structures next to each other, a comprehensive tree layout is created. When observing the layout broadly, each block acts as a super node, together forming a tree structure. This arrangement is similar to a d-ary heap, with super nodes serving as its parts. The value of 'd' in this structure corresponds to 'block_width × inter_child_count', where 'block_width' represents the number of leaf nodes in a block (nodes in the last layer of the block).
To determine the optimal parameter values, one must consider many different factors, such as the extra cost of computing the positions of the children of block leaf nodes as compared to regular nodes. Note that this ratio can also depend on the compiler optimization level. Therefore, to simplify this task, we measure the efficiency of a tree layout by the CPU time it takes to perform the heapsort algorithm on large arrays.

## Implementation

The data structure is represented by a single class named Par_Heap. During initialization, the constructor takes three main parameters that define the structure's layout. As shown in Figure 2, it calculates additional values such as 'block_size' and 'block_width' based on these parameters, which are essential for later operations.

```
Par_Heap(ll block_depth, ll intra_ch_cnt, ll inter_ch_cnt):
    block_depth(block_depth),
    intra_ch_cnt(intra_ch_cnt),
    inter_ch_cnt(inter_ch_cnt)
{
    block_width = 1;
    block_sz = 1;
    for (ll i = 0; i < block_depth; i++) {
        sub_block_sz += block_width;
        block_width *= intra_ch_cnt;
        block_sz += block_width;
    }
    block_ch_cnt = block_width*inter_ch_cnt;
}
```

*Figure 2: Par_Heap constructor, calculating the needed values based on the given parmeters*

In this research, we focused on implementing two essential methods required for the test: 'heapify' and 'heap_sort'. An important modification we made is in the 'heapify' method, which now accepts two arguments. One argument keeps track of the current super_node index, and the other, called 'localI', keeps track of the position within the current super_node.

Within the 'heapify' method, we initially determine if the current node is a block_leaf node. Based on this and the chosen layout during class construction, we calculate the positions of the node's children. The subsequent steps follow the typical heap behavior, selecting the node with the most of a property (such as smallness) and continuing the heapify process or terminating, similar to the traditional heap method. The complete 'heapify' implementation is detailed in Figure 3.

In cases where the 'inter_child_count' parameter is set to one, we utilized a simplified version of the code. In which we statically set the parameter to one and removed the 'for' loop used for finding the children of block_leaf nodes. This optimization reduced the number of CPU instructions required. However, a similar optimization for the 'inter_child_count' of 2 did not significantly improve our benchmarks. As a result, we discontinued this optimization for values greater than one.

```cpp
void heapify(ll I, ll localI) {
    // calculate the actual index of the current node in the data array
    ll index = I * block_sz + localI;
    // check if it's a block leaf node
    if (localI >= sub_block_sz) {
        // the current node is a leaf node
        // calculate the range of its childs
        ll itI = I*block_ch_cnt + 1 + (localI-sub_block_sz)*inter_ch_cnt;
        ll start  = (itI) * block_sz;
        ll end = min(start + inter_ch_cnt * block_sz, n);
        // iterate over the calculated range
        int mx = data[index];
        ll mx_I = I;
        for (; start < end; start += block_sz, ++itI) {
            // compare (set max)
            if (data[start] > mx) {
                mx = data[start];
                mx_I = itI;
            }
        }
        // terminate heapify if no child of the parent has a greater value
        if (mx_I == I) return;
        // swap & recursively call heapify
        swap(data[mx_I * block_sz], data[index]);
        heapify(mx_I, 0);
        return;
    }
    // if it's not a block leaf node, then the mapping to children is done
    // the same way as the regular d-ary heap:
    // calculate the range of its childs
    ll start = I*block_sz + localI * intra_ch_cnt + 1;
    ll end = min(start + intra_ch_cnt, n);
    // iterate over the calculated range
    int mx = data[index];
    ll mx_index = index;
    for (; start < end; ++start) {
        // compare (set max)
        if (data[start] > mx) {
            mx = data[start];
            mx_index = start;
        }
    }
    // terminate heapify if no child of the parent has a greater value
    if (mx_index == index) return;
    // swap & recursively call heapify
    swap(data[mx_index], data[index]);
    heapify(I, mx_index-I*block_sz);
}
```

*Figure 3: The complete implementation of the 'heapify' method used in Par_Heap data structure. Note that 'll' is the cpp 'long long' data type.*

## The search method:

The process of finding suitable Par-Heap parameters on any system is automated by a search method. In this study, we employed an exhaustive search approach due to the manageable search space. Notably, values of 'block_depth' greater than 1+log(N) (in base 'intra_child_count') yield an identical structure resembling a regular d'ary heap. This occurs because all nodes are confined within the first and only block (super_node). Additionally, our experiment involved setting manual limits on 'inter_child_count' and 'intra_child_count'.

The efficiency of a tree layout is measured by the CPU time required to perform in-place heapsort on an array of 80 million randomly generated integers. Table 1 displays CPU times for various tree layouts. Among these, layout (2, 9, 1) yielded the most favorable results. Consequently, this layout is employed in the comparison tests outlined in the upcoming Experiment section.

| Inter_child_count | Intra_child_count | Block_depth | | | | | | | | | |
|---|---|---|---|---|---|---|---|---|---|---|---|
| | | 1 | 2 | 3 | 4 | 5 | 6 | 7 | 8 | 9 | 10 |
| 1 | 2 | 5.58 | 3.34 | 2.67 | 2.39 | 2.3 | 2.38 | 2.56 | 2.87 | 3.29 | 3.7 |
| | 3 | 3.55 | 2.23 | 1.98 | 1.92 | 2.26 | 2.61 | 2.92 | 2.96 | 3.13 | 3.35 |
| | 4 | 2.92 | 1.91 | 1.81 | 2.06 | 2.39 | 2.6 | 2.68 | 2.87 | 3.08 | 3.2 |
| | 5 | 2.58 | 1.84 | 1.79 | 2.11 | 2.39 | 2.38 | 2.58 | 2.74 | 2.82 | 2.77 |
| | 6 | 2.38 | 1.73 | 1.84 | 2.16 | 2.19 | 2.36 | 2.54 | 2.62 | 2.57 | 2.26 |
| | 7 | 2.28 | 1.69 | 1.81 | 2.18 | 2.15 | 2.35 | 2.46 | 2.5 | 2.21 | 2.13 |
| | 8 | 2.18 | 1.66 | 1.85 | 2.13 | 2.16 | 2.36 | 2.47 | 2.37 | 2.09 | 2.08 |
| | 9 | 2.1 | 1.63 | 1.92 | 2.01 | 2.15 | 2.31 | 2.35 | 2.05 | 1.99 | 1.98 |
| | 10 | 2.08 | 1.64 | 1.95 | 1.97 | 2.16 | 2.3 | 2.27 | 1.96 | 1.98 | 2.01 |
| 2 | 2 | 3.43 | 2.74 | 2.45 | 2.3 | 2.41 | 2.52 | 2.86 | 3.12 | 3.47 | 3.83 |
| | 3 | 2.61 | 2.16 | 2.02 | 2.11 | 2.37 | 2.72 | 2.97 | 2.97 | 3.16 | 3.41 |
| | 4 | 2.43 | 1.97 | 1.9 | 2.2 | 2.54 | 2.59 | 2.73 | 2.95 | 3.17 | 3.3 |
| | 5 | 2.3 | 1.83 | 1.99 | 2.28 | 2.41 | 2.43 | 2.65 | 2.8 | 2.89 | 2.81 |
| | 6 | 2.2 | 1.8 | 2.02 | 2.33 | 2.21 | 2.42 | 2.6 | 2.68 | 2.63 | 2.29 |
| | 7 | 2.12 | 1.76 | 2.07 | 2.28 | 2.19 | 2.42 | 2.54 | 2.54 | 2.26 | 2.17 |
| | 8 | 2.08 | 1.79 | 2.11 | 2.16 | 2.23 | 2.44 | 2.54 | 2.41 | 2.12 | 2.09 |
| | 9 | 2.07 | 1.85 | 2.12 | 2.04 | 2.22 | 2.37 | 2.39 | 2.07 | 2 | 1.98 |
| | 10 | 2.04 | 1.89 | 2.16 | 2.02 | 2.23 | 2.36 | 2.3 | 1.96 | 1.97 | 2 |

*Table 1: cpu time (scale = 1e7 microseconds) needed for each layout to complete the heap sort operation on an array of size 80 million, filled with random 32bit integers*

## Experiment:

The experiment involved creating arrays (containing random 32bit integer values) of various sizes ranging from 10 to 1e8. We then applied the 'make_heap' operation followed by 'heap_sort' on this array. We compared two main heap data structures: our Par-Heap with semi-optimal parameters and the implicit binary heap. The comparison was based on the number of L2 and L3 total cache misses and the CPU execution time required to complete the task.

In our tests, the implicit binary heap was implemented in two ways: one utilizing standard C++ library functions, referred to as std-Heap, and the other method is a simpler less optimized implementation referred to as the standard-Heap. Each test was repeated 10 times, and the average results are presented in Figures 4,5,6. It is important to note that there might be some unavoidable noise in the number of cache misses reported by the PAPI library [10].

The findings indicate that our Par-Heap structure, with a semi-optimal layout, demonstrated significant improvements in both the number of cache misses and CPU execution time compared to the other two methods. All programs and tests were written in the C++ programming language and are accessible at our public github repository [9].

### Compilation optimization and compiler choice:

All methods discussed in this research were compiled for benchmarking purposes using the Clang compiler on O2 optimization level. Without the proper compiler optimizations, the overhead associated with the extra instructions required in our Par-Heap implementation increases and nullifies the improvements. Note that the focus in our implementation was on simplicity and readability of the code. For these reasons, a simple recursive implementation was chosen for the heapify method, leaving the optimization aspect entirely to the compiler.

### Test System Specifications:

The tests were conducted on a computer with the following specifications:

| | |
|---|---|
| CPU Model name | Intel(R) Core(TM) i7-7700HQ CPU @ 2.80GHz |
| L1 Data cache size | 128 KiB (4 instances) |
| L1 Instruction cache size | 128 KiB (4 instances) |
| L2 cache size | 1 MiB (4 instances) |
| L3 cache size | 6 MiB (1 instance) |
| Architecture | x86_64 |
| CPU op-mode(s) | 32-bit, 64-bit |
| Address sizes | 39 bits physical, 48 bits virtual |
| Operating System | Fedora Linux |

Benchmark Results:

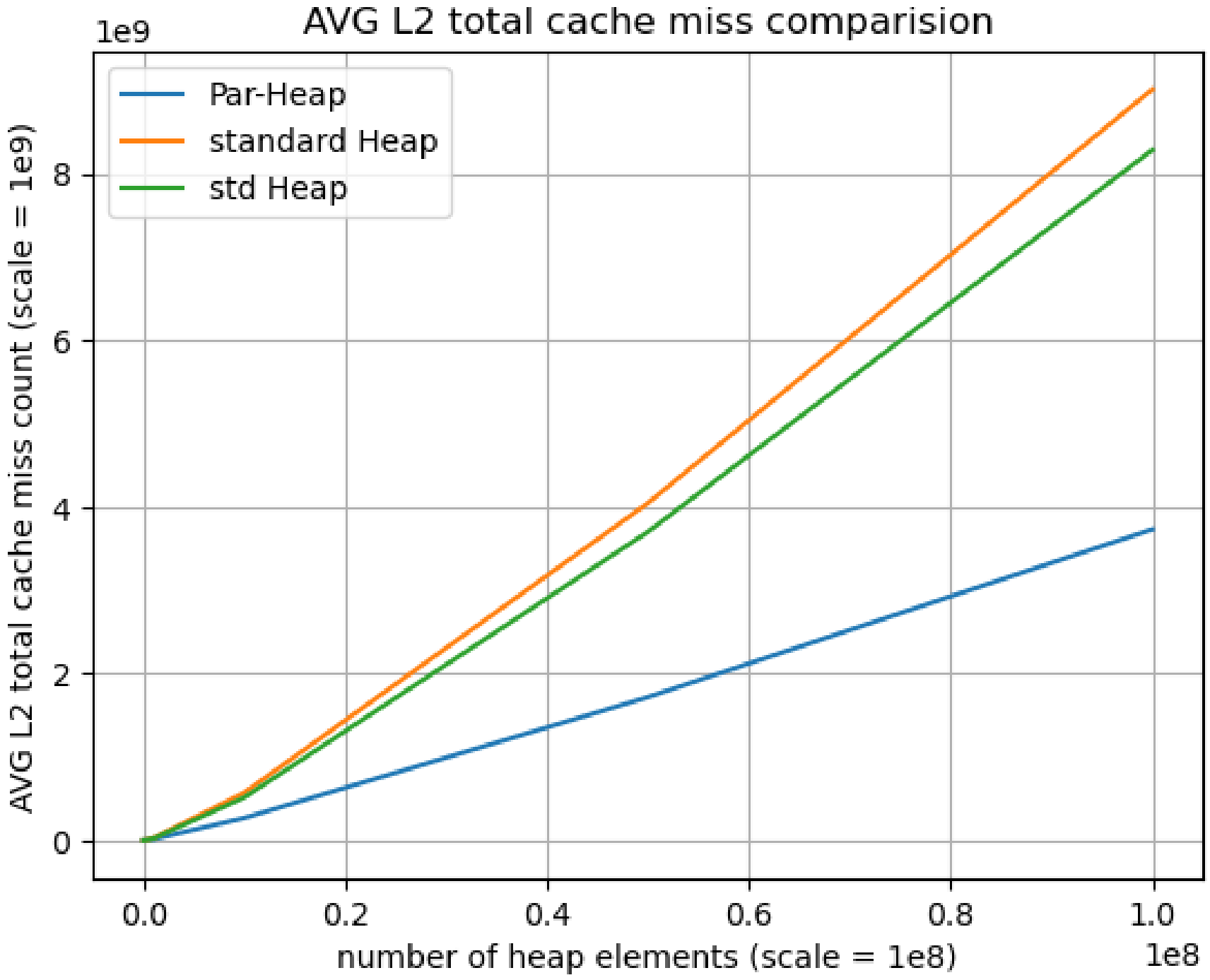

*Figure 4: Total L2 cache miss count when executing the heap-sort algorithm with varying numbers of heap elements.*

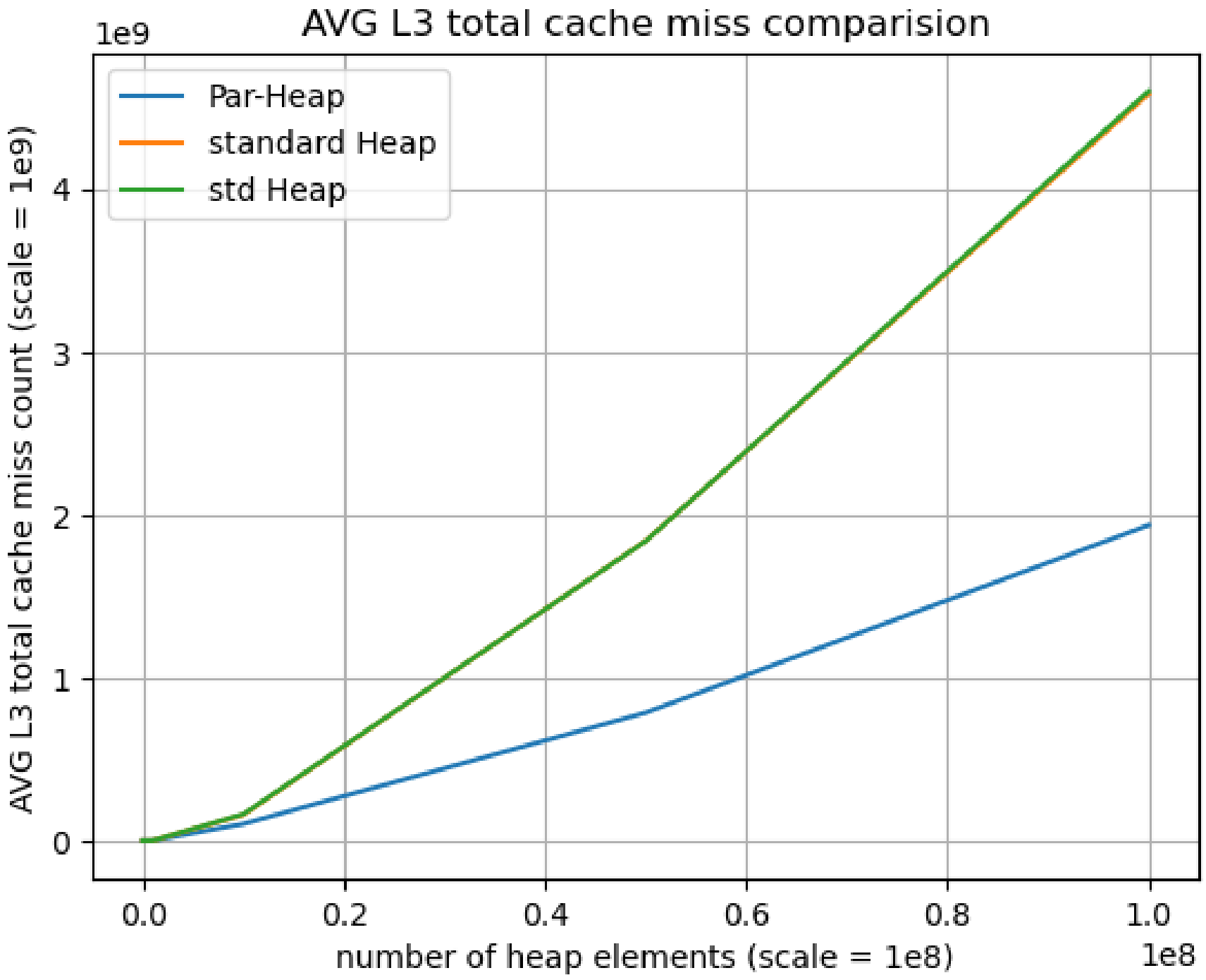

*Figure 5: Total L3 cache miss count when executing the heap-sort algorithm with varying numbers of heap elements. Notice the nearly identical results achieved by the 'standard' and 'std' methods.*

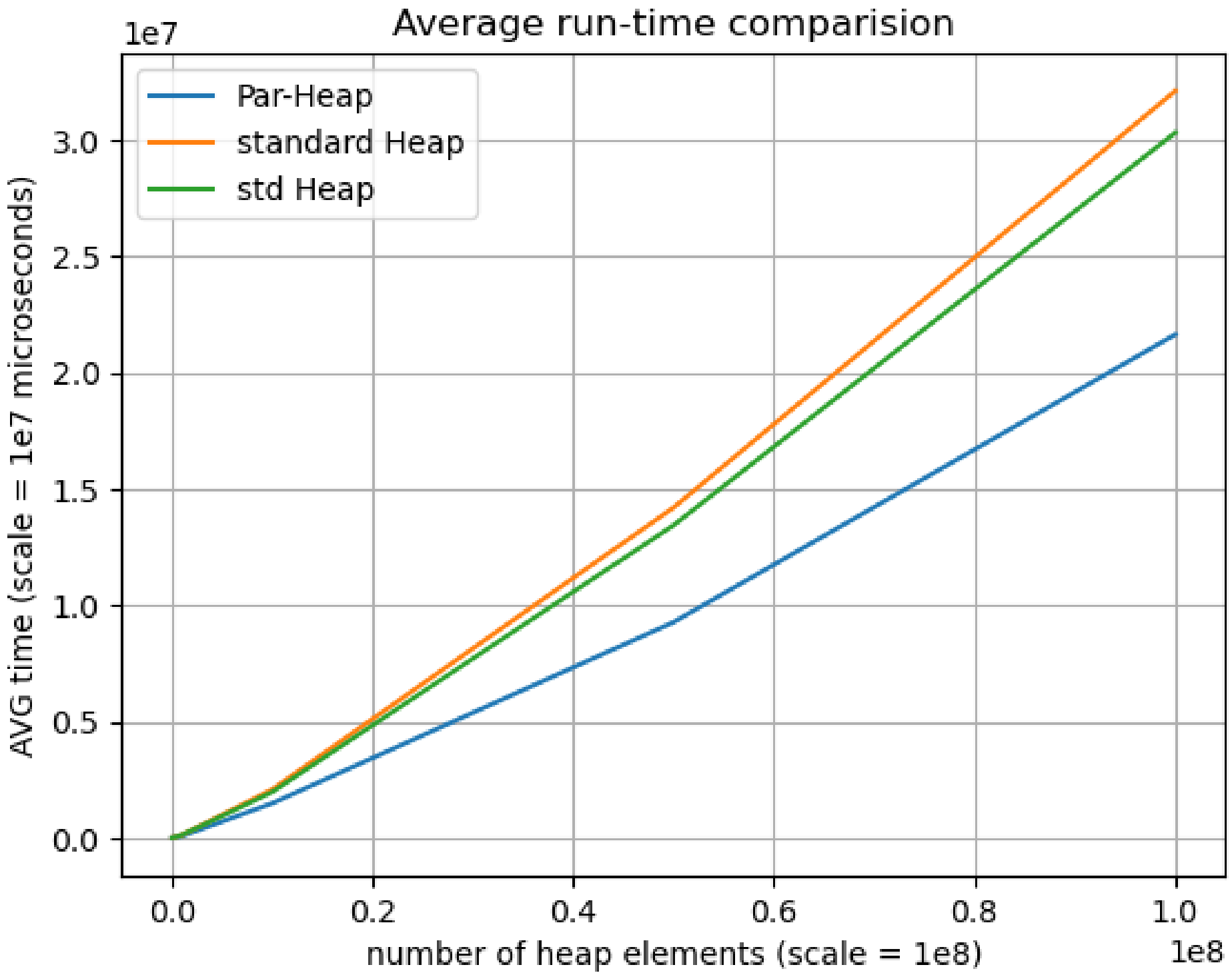

*Figure 6: Comparison of CPU Time spent executing the Heap-Sort algorithm using various methods discussed in this research.*

The Par-Heap data structure exhibits significant performance improvements compared to standard implicit binary heap implementations. Benchmarks demonstrate up to a 60% reduction in L3 cache misses, up to a 50% reduction in L2 cache misses, and up to a 32% reduction in execution time when employed with appropriate compiler optimizations. Note that a consistent clang compiler optimization level of O2 was employed for all methods in all benchmarks.

## Conclusion

This research has introduced a new approach to priority queues, focusing on optimizing the implicit tree layout and cache-friendliness. By incorporating adjustable parameters, we have developed a simple and flexible heap tree structure that can be tailored to diverse system conditions. Extensive testing has demonstrated that implementing the techniques outlined in this research can result in substantial performance improvements. Furthermore, our research emphasizes practicality by providing a simple and comprehensible implementation of the Par-Heap in this paper. Looking forward, future research endeavors could focus on improving the search method used for finding the optimal parameter values, aiming for a faster search algorithm.